\documentclass[]{emulateapj}

\def\ltsima{$\; \buildrel < \over \sim \;$}
\def\lsim{\lower.5ex\hbox{\ltsima}}
\def\gtsima{$\; \buildrel > \over \sim \;$}
\def\gsim{\lower.5ex\hbox{\gtsima}}
\def\be{\begin{equation}}
\def\ee{\end{equation}}
\def\no{\noindent}
\renewcommand{\vec}[1]{{\bf #1}}
\newcommand{\bhat}[1]{{\hat {\bf #1}}}

\begin{document}

\title{Hyper Velocity Stars \\and\\ The Restricted Parabolic 3-Body Problem}
\author{Re'em Sari$^{1,2}$, Shiho Kobayashi$^{3,1}$, Elena M. Rossi$^1$}
\affil{$^1$Racah Institute of Physics, Hebrew University, Jerusalem, Israel, 91904 \\}
\affil{$^2$Theoretical astrophysics 350-17, California Institute of
           Technology, Pasadena, CA, 91125 \\}
\affil{$^3$Astrophysics Research Institute, Liverpool John Moores University, United Kingdom}

\begin{abstract}
Motivated by detections of hypervelocity stars that may originate from
the Galactic Center, we revist the problem of a binary disruption by a
passage near a much more massive point mass. The six order of
magnitude mass ratio between the Galactic Center black hole and the
binary stars allows us to formulate the problem in the restricted
parabolic three-body approximation. In this framework, results can be
simply rescaled in terms of binary masses, its initial separation and
binary-to-black hole mass ratio. Consequently,
an advantage over the full three-body calculation is that
a much smaller set of simulations is needed to explore the relevant parameter space.
Contrary to previous claims, we show that, upon
binary disruption, the lighter star does not remain preferentially
bound to the black hole. In fact, it is ejected exactly in 50\%
of the cases.  Nonetheless, lighter objects have higher ejection
velocities, since the energy distribution is independent of mass. 
Focusing on the planar case, 
we provide the probability distributions for disruption of circular binaries and for the ejection energy.
We show that even binaries that penetrate deeply into the tidal
sphere of the black hole are not doomed to disruption, but survive in
$20\%$ of the cases. Nor do these deep encounters produce the highest
ejection energies, which are instead obtained for binaries arriving
to $0.1-0.5$ of the tidal radius in a prograde orbit. Interestingly,
such deep-reaching binaries separate widely after penetrating the tidal radius, but
always approach each other again on their way out from the black hole.
Finally, our analytic method allows us to account for a finite size of
the stars and recast the ejection energy in terms of a minimal
possible separation.  We find that, for a given minimal separation,
the ejection energy is relatively insensitive to the initial binary
separation.

\end{abstract}
\section{Introduction}

Hypervelocity Stars (HVSs) are stars with a velocity exceeding the
escape velocity of the Galaxy.  Currently, $16$ such stars have been
observed, 15 of which are thought to originate from our Galactic
Centre \citep{BGK+05,BGK+06,BGK+07,BGK+09,HHO+05} and one
from the Large Magellanic Cloud \citep{ENH+05}.  All have been
observed with radial velocities between 300~km/s  and 800~km/s
and almost all are located over 50~kpc away. For stars whose likely
origin is within the Galaxy and taking the galactic potential into account,
this translates to velocities of over 1000~km/s from the bulge.  The
observational strategy is such that most of the discovered HVSs
are faint B stars.  They have escape velocities from the surface of
the order of 600~km/s, well below the ejection velocity of 1000~km/s.
Thus, the standard mechanisms for producing
high velocity runaway stars, such as star-scattering and explosion as
a supernova of one component of a binary cannot work. A dynamical
interaction with a massive compact object is likely involved.

 In this paper, we adopt one of the leading models for the formation of
 HVSs: the breakup of a binary as it approaches the black hole in the
 Galactic Center \citep{Hil88}. Simple analytical arguments can be made to show the
 potential of this model to explain HVSs.
 
 If the binary of total mass $m$ has separation $a$, then tidal forces
from the black hole overcome the binary's mutual gravitational forces
at the tidal radius $r_t=a(M/m)^{1/3}$, where $M$ is the mass of the
black hole. The relative velocity of the binary components is of 
order $v_0=(Gm/a)^{1/2}$.  If the binary approached the black hole with
negligible energy, its center of mass moves at the tidal radius with
velocity of order $v_{\rm BH}=(GM/r_t)^{1/2}=v_0 (M/m)^{1/3}$ relative to the
black hole.
There are three ways to estimate the energies of the
individual components of the binary, assuming that they arrived with
negligible total energy. It is instructive to consider all three:
\begin{itemize}
\item {\bf Kinetic energy:} Adding or subtracting the relative velocity of the components, $v_0$, to the velocity around the black hole,
results in an additional kinetic energy of order $v_0v_{\rm BH} \sim v_0^2 (M/m)^{1/3}$. 
\item {\bf Gravitational potential energy:} The displacement of order $a$ in the position of each component of the binary, at a distance of
about $r_t$ from the black hole, results in a change in gravitational energy of $GMa/r_t^2 \sim v_0^2 (M/m)^{1/3}$.
\item {\bf Work:} The energy of each of the component in the black hole frame, is changing only due to mutual forces between the binary components. The force is of order $Gm/a^2$ and the length, 
in the black hole frame, over which it acts is $r_{\rm t}$. Therefore the work is 
$Gmr_t/a^2 \sim v_0^2(M/m)^{1/3}$. 
\end{itemize}
All these estimates provide an energy of order $v_0^2 (M/m)^{1/3}$.
If the binary dissolves, one component of the binary stays bound to
the black hole, the other escapes, with velocities at
infinity\footnote{We ignore the galactic potential in this paper.  By
``the velocity at infinity" we mean the velocity of the object once it
escaped the gravitational potential of the black hole, but did not yet
climb out of the potential of the rest of the galaxy.} of $v_0(M/m)^{1/6}$.
The encounter with the black hole therefore allowed for a larger
velocity by a factor of $(M/m)^{1/6}$ than the orbital velocity of the
binary.  For the parameters of our Galaxy and stars, $(M/m)^{1/6}\sim
10$, allowing ejections with thousands of kilometers per second.

 This is the theoretical framework in which \cite{Hil88}
 predicted the existence of HVSs.
 Later, it was discussed by \cite{YuT03} and \cite{GQ03}.
 After the observational discovery, many
 more papers on the subject appeared \citep{GPS05,GL06,GiL06,BKG+06,SHM07,PHA07,KBG+08},
 aiming to predict the properties of the ejected and/or captured stars.
 Other papers, \cite[e.g.][]{GL07,A09}, focused instead on the fate of binaries that are not 
 dissolved, but that in fact coalesce.
 The investigations so far have used three-body simulations or analytic methods that relied on results from
 three-body simulations. Consequently, only a limited set of parameters (e.g. for the binary mass ratio)
  have been explored.

In this paper,
we show  that this problem can be investigated with methods 
related to those used in the study of asteroids in the Solar System,
which exploit the enormous disparity in mass between the bodies involved.
Specifically, we can formulate it in terms
of a restricted 3 body problem, i.e. the motion of a single massless
particle under the influence of external time dependent forces. 
Our treatment is valid as long as the binary components are closer to each
other than to the black hole. Since $r_{\rm t}=a(M/m)^{1/3}$ our
approximation requires $(M/m)^{1/3} \gg 1$,
which is fully satisfied in the case of the black hole at the
Galactic Center and a binary of B stars, $(M/m)^{1/3} \cong 100$.
With the advantages of this analytic method, we can reach general conclusions
that do not depend on the physical properties of the system, such as masses and
binary semi-major axis. Moreover, the orbit integration is faster and more stable,
allowing us to handle more easily cases of close encounters between the bodies.

In \S \ref{sec:parabolic} we outline the formulation of the 3-body
problem in terms of a restricted parabolic problem.  In \S
\ref{sec:radial} we use the restricted radial problem to describes
binaries that penetrate deep into the tidal radius.  The radial
problem has a singularity at the time that the binary encounters the
black hole, and we use the results of the parabolic problem to pass
the singularity and continue to describe the evolution of the binary
on its trajectory away from the black hole.  In \S
\ref{sec:verification} we compare our results with 3-body numerical
integrations, and find excellent agreement.  We then use the parabolic
and radial formalism to investigate the probability of dissolving a circular, planar
binary and to obtain quantitative estimates of the ejection
velocities. We outline these results in \S \ref{sec:results}.

\section{The parabolic restricted 3 body problem} \label{sec:parabolic}
\subsection{The orbit}

In the case of interest $M/m \gg 1$ 
and the equation of motion for each of the binary members reads
\begin{equation}
\ddot{\vec r}_1=-{G M \over r_1^3}\vec r_1 + {Gm_2 \over |\vec r_1-\vec r_2|^3}  (\vec r_2- \vec r_1)
\end{equation}
\begin{equation}
\ddot{\vec r}_2=-{G M \over r_2^3}\vec r_2 - {Gm_1 \over |\vec r_1-\vec r_2|^3}  (\vec r_2- \vec r_1).
\end{equation}
where ${\vec r_1}$ and ${\vec r_2}$ are the respective distance from the black hole.
We therefore obtain that the equation for the distance between the two $\vec {\tilde r}\equiv \vec r_2-\vec r_1$ 
is
\begin{equation}
\label{vecr}
\ddot {\vec {\tilde r}}=-{G M \over r_2^3}\vec r_2+{G M \over r_1^3}\vec r_1 - {Gm \over {\tilde r}^3} \vec {\tilde r}.
\end{equation}
Now, we assume that the two masses are much closer to each other, and
to some effective parabolic trajectory $\vec r_{\rm m}$, than each of
them to the central mass $M$. As we already noticed in our Introduction,
this condition is actually enforced, up to the tidal radius, by the requirement of a large
mass ratio $M/m \gg 1$. Deep inside the tidal sphere, this assumption may break and we will discuss
the relevant conditions in \S\ref{sec:sigularity}.

We define the effective trajectory,
$\vec{r_{\rm m}}$, to be the parabolic orbit set by the position and
velocity of the center of mass of the binary when it is far away from
the black hole. The trajectory $r_{\rm m}$ is therefore described by
\begin{equation}
\label{r(f)}
r_{\rm m}={2r_p \over 1+ \cos f},
\end{equation}
where $r_{\rm p}$ is the distance of closest approach, and $f$ the angle from
the point of closest approach. The angle $f$, know as the true anomaly,
is a function of time, but one has analytically only the time as a
function of $f$: 
\begin{equation}
\label{t(f)}
\tilde{t} = {\sqrt{2 } \over 3} \sqrt{r_p^3 \over GM} \tan (f/2) \left(3  +  \tan^2 (f/2)  \right).
\end{equation}

\noindent
Note that the actual center of
mass will not generally move on this orbit, since the total force on
the binary does not equal the force that would act if the binary was a
single body. 

Linearizing the first
two terms of equation (\ref{vecr}) around $\vec r_{\rm m}$, we find that
the zero orders cancel out, and we obtain

\begin{equation}
\label{eq:rphys}
\ddot {\vec {\tilde r}}=-{G M \over r_{\rm m}^3}\vec {\tilde r}+3 {G M \over r_{\rm m}^5}(\vec r \vec r_{\rm m})\vec r_{\rm m}   - {Gm \over {\tilde r}^3} \vec {\tilde r}.
\label{eq:rdot2_p}
\end{equation}
 
Rescaling the distance between the bodies by $(m/M)^{1/3} r_{\rm p}$ and the time by $\sqrt{r_p^3/GM}$,
we can re-write eq.~(\ref{eq:rdot2_p}) in terms of the {\it dimensionless} variables\footnote{Later, for the radial problem (\S\ref{sec:radial}), we will rescale our variables differently, keeping the same symbols. Therefore, in the whole paper, $\vec{r}$ and $t$ should be understood as dimensionless, with a normalization that should be clear from the context.} $\vec{r}$ and $t$

\begin{equation}
\ddot {\vec r}=\left( r_p  \over r_{\rm m}\right)^3\left[ - \vec r+3 (\vec r \bhat r_{\rm m})\bhat r_{\rm m} \right]  - {\vec r \over r^3}.
\label{eq:r_vec}
\end{equation}

Since $\bhat r_{\rm m}=(\cos f, \sin f, 0)$, $r_{\rm m}/r_{\rm p}=2/(1+\cos
f)$ and we set $\vec{r}=(x,y,z)$, explicit equations in terms of
dimensionless Cartesian coordinates read: 
\begin{eqnarray}
\label{ddotx}
\ddot x={(1+\cos f)^3\over 8} \left[ - x+3 (x \cos f+y \sin f )\cos f   \right] \cr
- {x \over (x^2+y^2+z^2)^{3/2} }\,,
\end{eqnarray}
\begin{eqnarray}
\label{ddoty}
\ddot y={(1+\cos f)^3\over 8} \left[ -y+3 (x \cos f+y \sin f )\sin f
			     \right] \cr
  - {y \over (x^2+y^2+z^2)^{3/2} }\,,
\end{eqnarray}
 \begin{equation}
\label{ddotz}
\ddot z=-{(1+\cos f)^3\over 8} z   - {z \over (x^2+y^2+z^2)^{3/2} }\,,
\end{equation}
\begin{equation}
\label{tbar(f)}
t ={\sqrt 2 \over 3} \tan (f/2) \left(3  +  \tan^2 (f/2)  \right).
\end{equation}
Equations (\ref{ddotx})-(\ref{ddotz}) are the equivalent of the Hill
equations \citep{Hill1886}, for the parabolic rather than circular problem. 
Since equation (\ref{tbar(f)}) is implicit, for numerical applications
it may be preferable to use its differential form 
\begin{equation}
\label{fdot(f)}
\dot f=\sqrt{2} (1+\cos f )^2/4 .
\end{equation}

\subsection{Free Solutions}
Just like the Hill equations, equations~(\ref{ddotx})-(\ref{ddotz})
have free solutions, those that ignore 
the interaction term $\vec r/r^3$. 
These can be found mathematically, but physical intuition
facilitates a swift solution. Since we have a set of three linear
differential equations of the second order, all solutions are a linear
combination of six independent solutions.  Each could be physically
obtained by taking the difference between an orbit infinitesimally
close to a parabolic orbit and the parabolic orbit itself.
We list the solutions below, stating which orbital element has been varied.
\begin{itemize}
\item Variation in the argument of periapsis
\begin{equation}
\label{varomega}
x=-{\sin f \over 1+\cos f} \quad , \quad y={\cos f \over 1+\cos f} .
\end{equation}
\item Variation in the time of periapsis
\begin{equation}
\label{varperi}
x=-\sin f \quad , \quad y=1+\cos f .
\end{equation}
\item Variation in the periapsis distance
\begin{equation}
\label{varr_p}
x= 2-\cos f  \quad , \quad y={- \cos f  \tan (f/2)} .
\end{equation}
\item Variation in the eccentricity at fixed periapsis
\begin{equation}
\label{varecc}
x=\left( 8+12\cos f \right) \tan^4(f/2),
\end{equation}
\[
 y={35 \sin f-2 \sin (2f) +3 \sin (3f) \over  (1+\cos f)^2} .
 \]
\item Rotation around the apsidal line
\begin{equation}
\label{varrotaps}
x=y=0 \quad , \quad z={2 \sin f \over 1+\cos f} .
\end{equation}
\item Rotation around the latus rectum
\begin{equation}
\label{varrectum}
x=y=0 \quad , \quad z={2 \cos f \over 1+\cos f}\, .
\end{equation}
\end{itemize}
In the above expressions, $f$ is a function of $t$ as given by
equation~(\ref{tbar(f)}). The first four solutions are planar, i.e.
$z=0$, while the last two solutions are one dimensional, $x=y=0$. 
It is easy to check the validity of these expressions by substituting
them into equations~(\ref{ddotx})-(\ref{ddotz}), and using eq.~(\ref{fdot(f)}).

\subsection{The Energy}
 We are ultimately interested in the fate of a star in a binary, following
its encounter with the black hole. If the binary -- approaching the hole on a parabolic orbit-- is torn apart, 
a star can either become bound to the black hole or be ejected from the
system. To distinguish between these two possibilities, we calculate
its energy as a function of time, including the negative
gravitational energy due to the black hole. 
Initially, at large distances, the specific energy of one member is simply $\sim - v_{0}^2$.
After the binary disruption, the analytical arguments in our Introduction suggest that 
its energy is larger by a factor of $(M/m)^{1/3} \gg 1$.
We thus neglect the term due to the self-gravity of the binary.
In addition, for $\left(M/m\right) \gg 1$, the change in energy of the massive black hole can also be neglected
 and the total energy of $m_1$ reads

\begin{equation}
E_1=-{GMm_1 \over r_1}+m_1 \left| \dot {\vec r}_1\right|^2/2 \,.
\end{equation}
To zero order, we can replace $\vec r_1$ by $\vec r_{\rm m}$, but since $r_{\rm m}$ is a parabolic orbit that zero order
energy vanishes. The first order terms are
\begin{equation}
\label{E1gen}
E_1={GMm_1 \over r_{\rm m}^2}(r_1-r_{\rm m})+ 
m_1 \dot {\vec r}_{\rm m} ( \dot {\vec r}_1- \dot {\vec r}_{\rm m}),
\end{equation}
or, using our rescaled variables,
\begin{equation}
E_1=-{GMm_1m_2 \over m r_p}\left(m\over M\right)^{1/3} \left[  {r_p^2 \over r_{\rm m}^2}\vec r\bhat r_{\rm m}+{\dot {\vec r}_{\rm m}  \over r_p} \dot {\vec r}\right].
\label{eq:E1}
\end{equation}

\no
Since in our limit  the total energy of the system is zero, the energy of the other body, $m_2$, is simply $E_2=-E_1$.
In the following, it is useful to define the
penetration factor $D = r_{\rm p} /r_{\rm t}$.
In terms of our dimensionless Cartesian coordinates eq~(\ref{eq:E1}) is given by
\begin{equation}
\label{E1par}
E_1  = -E_2 =  - {G m_1m_2 \over a\, D} \left(M\over m\right)^{1/3} \times
\end{equation}
\[
\times \left[  {(1+\cos f)^2 \over 4}(x\cos f+y\sin f) + \frac{-\sin f \dot x +(1+\cos f) \dot y }{\sqrt 2}\right].
\]

If the binary dissolves, this energy tends to a constant, since the 
body is eventually moving only under the conservative force of the
black hole. Mathematically, this means that the 
first term on the right-hand side of eq.~(\ref{eq:r_vec}) now dominates, and  
the problem is linear. 

The negative of a solution is thus a solution.
 But the energy is {\it also} linear in the coordinates.
Therefore, a body starting with $\pi$ phase
difference will have the same final energy in absolute value but opposite in sign.
This is independent of the mass of the star.
The important consequence is that, of
the disrupted cases, half would have the heavier object bound and the
lighter escaped, while half would have the opposite.
These findings are at odds with those of Bromley
et al (2006) who find that, for large $m_1/m_2$ ratios, the lighter object usually becomes bound.

We also note that, to this lowest order, the $z$ component of the binary motion
has no effect on the energy, thus it does not determine whether a body would be ejected. 

Finally, we explicitly write the energy,
 when we can neglect the interaction term between the two stars.
The solution in this ``free'' regime\footnote{Even in the non free
regime, where the gravitational forces between the two stars are
important, one can expand the solutions in terms of the free
solutions, except that the coefficients will be time dependent. Our
expression of the energy will still be given by $B$, except that $B$
now is time dependent.}, is a linear combination of
equations~(\ref{varomega})-(\ref{varrectum}).
For each free solution the energy is a constant of motion.
Therefore the total energy
is a linear combination of these constants.
However, any solution constructed from two bodies in infinitesimally
close parabolic orbits has zero energy. Only
eq.~(\ref{varecc}), that describes the relative orbits of two bodies where one
with $e\ne 1$, gives a finite constant energy when substituted into
eq.~(\ref{E1par}),

\begin{equation}
\label{E1B}
E_1=-E_2=- {G m_1m_2 \over a\, D} \left(M\over m\right)^{1/3}20 \;B,
\end{equation}
where $B$ is the coefficient of eq.~(\ref{varecc}) in the linear
expansion. 

\section{Deep Penetrators and the radial restricted 3 body problem}\label{sec:radial}

The previous section allowed for arbitrary penetration factors $D$.
The presence of the factor $1/D$ in eq.~(\ref{E1B}) seems to suggest that for a given initial star separation $a$
arbitrary large energies can be attained for binary trajectories penetrating 
far into the tidal disruption region, ($D \ll 1$). However, as we show here,
this is not the case.
We proceed to investigate this case of deep penetrators by considering 
 a different, even simpler, limit of the equations.
 The periapsis distance of the center of mass, $r_{\rm p}$, becomes
now irrelevant and instead the center of mass can be taken to move on a
radial orbit given by
\begin{equation}
r_{\rm m}=\left( 9GM {\tilde t}^2\over 2 \right)^{1/3}.
\end{equation}
We can now scale the distance to the initial semi-major axis of the binary, $a$, and the time to
the inverse angular frequency of the binary with that semi-major axis $\sqrt{Gm/a^3}$.
Taking the direction of the binary center of mass with respect to the BH to be $\bhat r_{\rm m}$, equation (\ref{eq:rphys}) simplifies to
\begin{equation}
\ddot {\vec r}=\left( 2  \over 9t^2\right)\left[ - \vec r+3 (\vec r \bhat r_{\rm m})\bhat r_{\rm m} \right]  - {\vec r \over r^3}.
\end{equation}
In Cartesian coordinate, where $\bhat x$ is the direction of the binary's center of mass, we have:

\begin{equation}
\label{rddotx}
\ddot x={4\over 9 t^2} x   - {x \over (x^2+y^2+z^2)^{3/2} }\,,
\end{equation}
\begin{equation}
\label{rddoty}
\ddot y=-{2\over 9t^2} y - {y \over (x^2+y^2+z^2)^{3/2} }\,,
\end{equation}
 \begin{equation}
\label{rddotz}
\ddot z=-{2 \over 9t^2 } z   - {z \over (x^2+y^2+z^2)^{3/2} }\,,
\end{equation}

The energy, as given by equation (\ref{E1gen}) is
\begin{equation}
\label{E1rad}
E_1=-E_2=-{Gm_1m_2 \over a} \left(M \over m\right)^{1/3} \left(2 \over 9 \right)^{2/3} |t|^{-4/3} \left( x +3 t \dot x  \right).
\end{equation}

\subsection{Special Solutions to the radial restricted problem}

\subsubsection{Homogeneous Collapse}
It is simple to verify that the following 
\begin{equation}
x(t)=\pm \left( \frac{3}{2}\right)^{1/3} |t|^{2/3} \quad , 
\quad y(t)=z(t)=0,
\end{equation}
is a solution to our equations.
This solution is the analog of a homogenous collapse. In these
solutions, the small masses $m_1$ and $m_2$ accelerate towards each
other, by a combination of tidal forces and mutual acceleration, at
the same rate that their center of mass is accelerating towards the
central mass. These solutions, just like the stationary solution at the
$L_1$ and $L_2$ points in the circular restricted problem, and just like the
homogeneous collapse of dust, are unstable for deviations in the x-direction.

\subsubsection{Free Solutions}
In the limit of large separation between the small masses or close to $t=0$ where the distance to the black hole is small,
tides dominate over the mutual gravity, and the problem become separable in $x$, $y$ and $z$.
It admits  the following solutions:
\begin{eqnarray}
\label{freerad}
x(t) & = & A_x |t|^{-1/3}+ B_x |t|^{4/3}, \cr
y(t) & = & A_y |t|^{1/3}+ B_y  |t|^{2/3},\cr
z(t) & = & A_z |t|^{1/3}+ B_z  |t|^{2/3} .
\end{eqnarray}
These are the equivalent of the free solutions given for the
parabolic restricted problem in equations
(\ref{varomega})-(\ref{varrectum}).  Just like those, they can be
given physical interpretations. The $A_x$ solution describes two
particles that have the same trajectory, but are slightly separated in
time. The $B_y$ and $B_z$ solutions, describe particles going on
slightly different radial paths, each of zero energy. The $B_x$
solution describes the relative orbits of two particle going on the
same radial path, but with slightly different energies.  The $A_y$ and
$A_z$ describe the relative orbits of particles with slightly
different angular momentum.  

Note, that both exponents of the $y$ and
$z$ terms are positive, while one exponent of the $x$ term is
negative.  This means that at times close to zero, the particles are
very close in $y$ and $z$ but are separated in $x$: $y(t \rightarrow -0)=z(t
\rightarrow -0)=0$ while $x(t \rightarrow -0)=\pm \infty$. 

As in the parabolic case, one of the coefficients of the solution is related to energy. 
Substituting the equation for the x component and its derivative into eq.~(\ref{E1rad}) we get, 
\begin{equation}
\label{E1Bx}
E_1={{Gm_1m_2 \over a} \left(M \over m\right)^{1/3} } 5 \, (2/9)^{2/3} \, B_x .
\end{equation}

This can be regarded has the equivalent of eq. (\ref{E1B}) in the limit of $D \rightarrow 0$.
It is evident, that the energy in this limit is finite, and the divergence suggested by 
eq. (\ref{E1B}) is not real. Moreover, as we will show later, the highest ejection velocities
are {\it not} obtained in this limit. 

\subsection{The $t=0$ singularity}
\label{sec:sigularity}

From the discussion above, we learned that, beside a set of zero
measure of initial conditions, the binary components 
acquire an increasingly large separation as they approach the large mass $M$. It is
therefore tempting to conclude that for deep penetrators, binaries
always dissolve. In fact, the two components of the binary approach each other
again at later (positive) times.
Thus, to know the final outcome of the scattering, we should overcome the singularity and 
 follow the orbits of the light bodies beyond the time of the encounter with the black hole.
 
The difficulty arises from the assumption of a {\it purely} radial orbit for the binary center of mass.
A {\it deep} ($D \ll 1$) parabolic orbit would parallel closely the radial one and get 
around the black hole smoothly.  Once across the singular region, the radial equations 
are valid and the integration of the orbit can be resumed.
We further notice that when the binary is well within the tidal
radius, $|t| \ll  1$ the mutual gravity of its members can 
be neglected and their separation follows a free solution.
Therefore,  the resolution of our problem ultimately lies in finding which 
free parabolic orbit  reduces --in the deep penetration approximation--
 to the free radial orbit that an approaching binary is following, while still farther away from its periapsis, 
$|t| \gg D^{3/2}$.

This translates in expanding eqs.~(\ref{varomega})-(\ref{varrectum})
around $f= -\pi$ and comparing them with eqs.~(\ref{freerad}).
Interestingly, we find that there is a one to one correspondence: solutions with
coefficients $A_x$, $A_y$ and $A_z$ correspond to eq.~(\ref{varperi}),
eq.~(\ref{varr_p}) and eq.~(\ref{varrotaps}) respectively.  Likewise, the
solutions with $B_x$, $B_y$ and $B_z$ correspond to eq~(\ref{varecc}),
eq.~(\ref{varomega}) and eq.~(\ref{varrectum}) respectively.  
This is as expected given their physical interpretation. 
To make an example, we expand eq.~(\ref{tbar(f)}) and the $x$ component
of eq.~(\ref{varecc}) multiplied by some coefficient $B$ around $f=-\pi$. We obtain
\be
x=-6^{4/3}\, |t|^{4/3}B.
\ee
This is the behavior of the  radial
solution with coefficient $B_x$. In fact, comparing the two solutions for $x$
and accounting for the different spatial and temporal dimensions, we get
\be
B_x=-6^{4/3} \, D^{-1}\,B.
\ee
With this relation, the two equations (\ref{E1B}) and (\ref{E1Bx}) are equivalent as we expect.

From this comparison, we also learn the behaviour of the binary separation
around $t=0$. There, the $A_x |t|^{-1/3}$ term dominates, corresponding to
the free parabolic solution derived by slightly varying the time at periapsis (eq.~\ref{varperi}). 
The latter, however, does not diverge at $t=0$. Instead, it describes a circle centered around $x=0$ with $y=R$, where $R$ is the radius of the circle. 

A key observation for our problem is that all parabolic orbits 
which reduce to the $A_{\rm s}$ ($B_{\rm s}$) solutions are asymmetric (symmetric)
function of time where the subscript ${\rm s}$ stand for $x,y$ or $z$. We conclude that we can ferry a free solution 
across $t=0$, from negative to positive times, by simply changing the sign of its $As$ coefficients.

A concern might arise from the fact that our assumption of small binary separation
relative to the black hole may break down at times where the binary is close to its periapsis.
Indeed, at small enough times, the binary separation is about
$a|t|^{-1/3}$ 
while the distance to the black hole  decreases as
$r_{\rm m} \sim a\,(M/m)^{1/3}|t|^{2/3}$. They match at $|t| = (m/M)^{1/3}$, from which follows that,
if the binary gets within a distance of $r_{\rm m}/a < (m/M)^{1/9}$ from the black hole,
our approximation is no longer valid. This occurs for a
penetration factor for the orbit smaller than $D<(m/M)^{2/9}$. When this
happens, also the parabolic formalism incorrectly traces the trajectory
for $|t|< (m/M)^{1/3}$. 
Specifically the true orbit deviates from the circle given by  eq.~(\ref{varperi}).  
Yet, the rest of our conclusions, including the energy of the
particles will not be affected. Using the same work
argument in our introduction, one derives that the energy gained during the periapse passage is
smaller by a factor of $D^2$  than that gained around the tidal radius. 
One factor of $D$ come from the smaller distance around the black hole,
and the other from the larger distance between the stars resulting in a smaller
force between them. We conclude that -- for any impact parameter--
both energy and orbit obtained with the method described above
are fairly accurate for $|t| >(m/M)^{1/3}$.


\section{Numerical Verifications}\label{sec:verification} 
We test our approximated equations against three-body simulations
of a binary evolving around a much more massive black hole.

Both the three-body code and the code that numerically integrates our equations
are provided with a 4th-order Runge-Kutta integration scheme. In the three-body code 
the binary center of mass is initially either on a parabolic (\S \ref{subsec:para})
or on a radial (\S \ref{subsec:radial}) orbit. The binary consists of a primary star with mass $m_1$ and of a
secondary with $m_2=m_1/3$. In the examples shown in this section, the
black hole-binary mass ratio is set to be $M/m=10^6$.  The binary's
orbit is assumed to be initially circular in the comoving frame of the
binary center of mass.  The initial configuration of the system is charactered by three
parameters. First, the initial distance of the binary center of mass to the black hole,
$r_0$.  However, as long as a simulation starts at a large enough
radius, $r_0\gg r_{\rm t}$, the orbits are largely independent of it. 
In our runs we assume $r_0=10 r_t$, which is sufficient for convergency.
Second,
the initial (at $t_0=t(r_0) <0$) binary phase, $\phi_0$. We
parameterize it using the {\it effective} phase $\phi$ at $t=0$ (i.e. at the
periapsis passage for a parabolic orbit or at $r=0$ for a radial one),

$$\phi_0=S\,\omega t_0 +\phi,$$  where $\omega$ is the (constant)
angular velocity of the binary at $r\gg r_{\rm t}$. Naturally, the {\it actual} phase at $t=0$ is in general
different from $\phi$, due to of the black hole tidal forces.
All angles are
measured from the x-axis. Finally, we should specify the direction of
rotation of the binary $S$, as viewed in the non-rotating frame: i.e.
the relative orientation of the the angular momentum of the binary
around the black hole and of a star around the binary center of mass. For a planar orbit
there are two possibilities: the angular momenta are aligned, $S=1$,
in which case we call the orbit {\it prograde}, or they are
anti-aligned, $S=-1$, and we call it a {\it retrograde} orbit. In
addition, for the parabolic case, the system has a fourth
parameter: the penetration factor $D$.

\subsection{Parabolic Orbits}
\label{subsec:para}
\begin{figure}
\epsscale{1.15}
\plotone{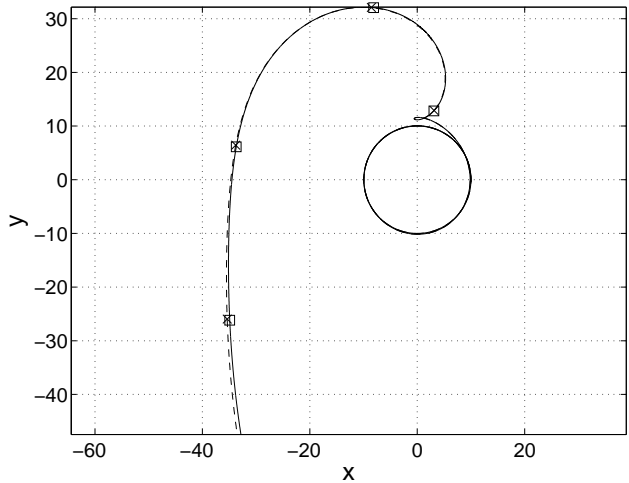}
\caption{The orbit of the secondary star in the primary comoving frame. 
 The binary center of mass is initially on a parabolic trajectory. We assume a prograde binary orbit,
 $M/m=10^6, m_1/m_2=3, D=0.1, \phi=4\pi/5$, and $r_0=10 r_t$.
 The three-body solution
 (solid line) is compared with our approximated solution (dashed line). 
 Squared (three-body) and cross (our solution) marks show the position of the secondary at
 $t=-3,0,3$ and $15$. Lengths and time are in units of $(m/M)^{1/3}r_p$ and
 $(r_p^3/GM)^{1/2}$, respectively. 
\label{fig:orbits1}}
\end{figure}

In this section, we compare the evolution of the binary stars obtained with
the three-body code and our parabolic formalism (eqs.~\ref{ddotx}-\ref{tbar(f)}).
We assume a prograde orbit for the binary, $D=0.1$ and $\phi=4\pi/5$.
The result is shown in Figure~\ref{fig:orbits1}, where
we plot the orbit of the secondary star in the comoving frame of the primary. 
After disruption around periapsis, the secondary star is captured by
the black hole on an elliptical orbit, while the primary is ejected
from the system.
Clearly, the three-body curve is accurately reproduced by our set of approximated equations.
The energy, for example, differs at a $0.1\%$ level after the binary disruption.

In addition, we compared the performance of the two
numerical methods, for $M/m \gg 1$. The calculation of our approximated solution is faster for two reasons.
First, for forth order scheme, numerical convergence is achieved with a time-step  about $\sim (M/m)^{1/12}$ times smaller.
Second, we integrate seven equations, instead of eighteen, since we only follow the binary
stars relative distance. This moderate speed-up, combined
with the fact that most dependencies are
analytic, allows us to explore more easily a wider
portion of the parameter space.

\subsection{Radial Orbits}
\label{subsec:radial}
\begin{figure}
\epsscale{1.15}
\plotone{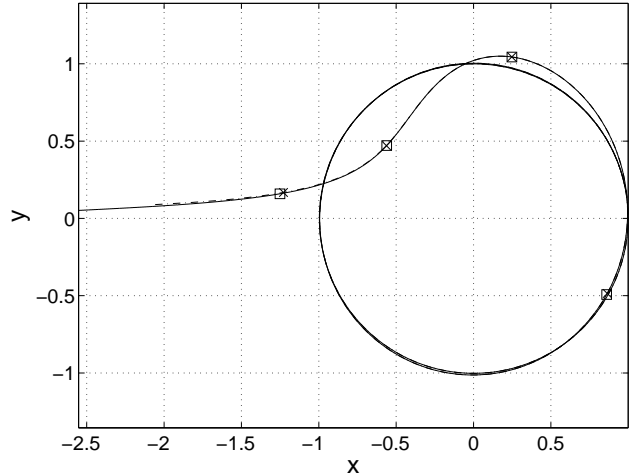}
\caption{As Figure~\ref{fig:orbits1}, but for an initial 
radial orbit for the binary center of mass. All relevant parameters are the same. 
Here, the position of the secondary is marked at $t=-10^{-2},-10^{-1},-1$ and $-3$.
Length and time are in unit of $a$ and $\sqrt{a^3/Gm}$, respectively.
 \label{fig:orbits2}}
\end{figure}

We now consider a binary moving on a radial orbit of zero energy 
and test the accuracy of our radial approximation
(eqs.~\ref{rddotx}-\ref{rddotz}). 
The trajectory of the separation
between the stars is shown in Figure~\ref{fig:orbits2}, for a binary
with a phase $\phi=4\pi/5$.  
Here, we plot the orbit up to $t=-2\times10^{-3}$.
The assumption of a relative small star separation is expected to
become invalid around $|t| \sim (m/M)^{1/3} = 10^{-2}$.
Again, there is a good agreement between our solution 
and the three-body calculation.
Comparing the two terms on the right-hand side
of eq.~(\ref{rddotx}), one expects that the tidal force 
dominates for $|t| \ll  1$. Indeed, we find numerically that the
deviation from the initially circular orbit becomes 
significant around $|t|\sim 1$.  For $|t|\ll 1$, the orbit 
approaches the free solutions 
$x \propto  |t|^{-1/3}$ and $y \propto |t|^{1/3}$. 
Finally, we also reproduce at a percentage level the energy evolution as a function of time.
For the specific example in Figure~\ref{fig:orbits2}, 
we get an energy that differs only by half a percent at $t\sim 0$ from the three-body result.

\subsection{Deep Penetrators}
\begin{figure}
\epsscale{1.2}
\plotone{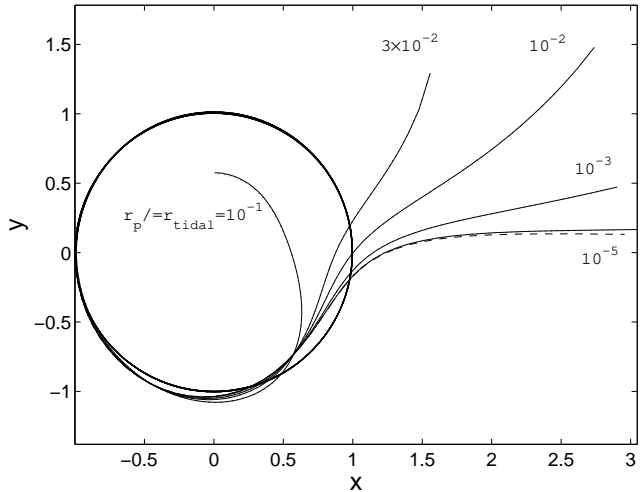}
\caption{As Figure~\ref{fig:orbits1}, but for different penetration factors
as labelled. The binary's orbit is prograde with parameters $M/m=10^6, \phi=0, r_0=10\, r_{\rm t}$.
Lengths are in unit of the initial binary separation.
The parabolic approximation (solid curves) and the three-body
 results (dotted  curves) are pratically indistinguishable.
In addition, we plot the radial orbit (dashed curve) for the same parameters.
\label{fig:orbits3}}
\end{figure}

In the previous sections, we have established that both our sets of 
equations well reproduce the full three-body calculation.
We now investigate their relationship. In particular, we show 
how a parabolic solution reduces to a radial one, in the limit 
in which a binary plunges deeply into the tidal sphere of influence of
the black hole. At this aim, we employ our parabolic formalism to
 numerically calculate binary orbits with increasingly smaller penetration factors,
(Figure~\ref{fig:orbits3}, solid lines). In this example, the orbit degenerate to a
radial one (dashed line) for $D \sim 10^{-5}$.

\section{Results}\label{sec:results}

We are now in the position to statistically describe the
properties of a binary after its encounter with a black hole. 
We focus here on the planar case.
For given masses of the three-bodies and semi-major-axis of the binary stars,
the fate and the final energy of the binary members depend only on the
penetration factor $D=r_{\rm p} / r_{\rm t}$ and on the effective binary
phase $\phi$. 

\subsection{ The Fate of the binary}
\begin{figure}
\epsscale{1.2}
\plotone{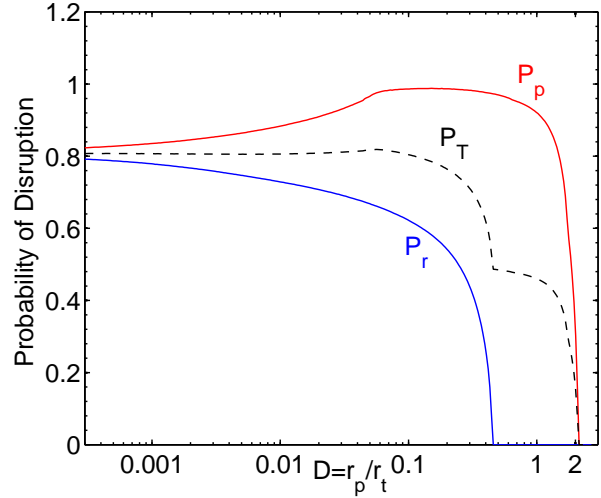}
\caption{Probability of disruption as a function of the penetration
factor $D$, for prograde orbit 
($P_{\rm p}$, red solid line), retrograde orbit ($P_{\rm r}$, blue solid line) 
and total $P=(P_{\rm p}+P_{\rm r})/2$. 
The binary phase $[0 ~\pi]$ is sampled with 3000 equally spaced grid points
for each penetration factor.}
\label{fig:dis_f}
\end{figure}

Contrary to naive expectations, we find that for $D \ll 1 $
a non-negligible fraction of the binaries are {\em not} disrupted (Fig.~\ref{fig:dis_f}).
For $D < 10^{-1}$, the fraction of
disrupted cases saturates at a
level of $\sim 80\%$. The best chance of disruption is for a binary in
prograde orbit, with $D = 0.15$. 
A comparison between the curves for prograde (labelled $P_{\rm p}$) and 
retrograde ($P_{\rm r}$) orbits underlines the well known fact that
retrograde binaries tend to be more stable against tidal disruption.
Nevertheless, the two curves converges for $D \ll 1$. 
This is because in this limit the binary center of mass approaches
the black hole in an almost radial fashion, so that the angular momentum
of the binary around the black hole is very close to zero. Indeed, in
the radial formalism, there is no distinction between
prograde and retrograde orbits. Consistently, when we calculate the
disruption probability for a binary on a radial orbit, we find a
survival fraction of $19\%$.
Interestingly, we observe that binaries that avoid disruption tend
to tighten, with their final semi-major axis $a_{\rm f}<a$. 
For instance, for radial orbits, the encounter with the black hole
produces harder binaries in about $90\%$ of the cases.

\subsection{Final energy of ejected stars }

\begin{figure}
\epsscale{1.25}\plotone{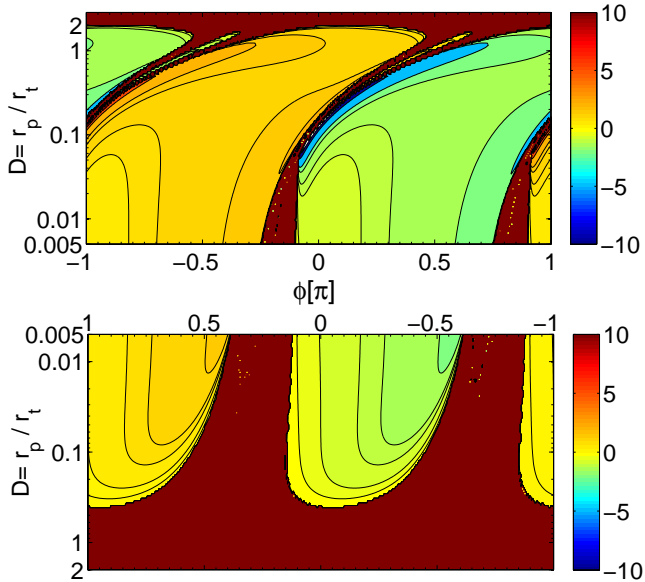}
\caption{Final energy contour plots, in the phase-penetration factor 
 plane for a given initial separation $a$. {\it upper panel:}
 prograde orbits. {\it lower panel} retrograde orbits.  
Energy is in unit of $(Gm_1m_2/a)(M/m)^{1/3}$ and the phase in unit of $\pi$.
In the lower panel axis are reversed. The energies have
 been computed using the restricted parabolic formalism. The figure is
 constructed with a grid of 400 $\phi$ values (equally spaced linearly) 
and 100 $D$ values (equally spaced logarithmically). We only plot the final
 energy of a member of a {\it disrupted} binary, which can be either
 captured by the black hole $E < 0$, or ejected $E>0$. 
The flat amaranth (dark red) region shows the region in 
the parameter space where the binary survives disruption.
\label{fig:ene_contours}}
\end{figure}

\begin{figure}
\epsscale{1.15}
\plotone{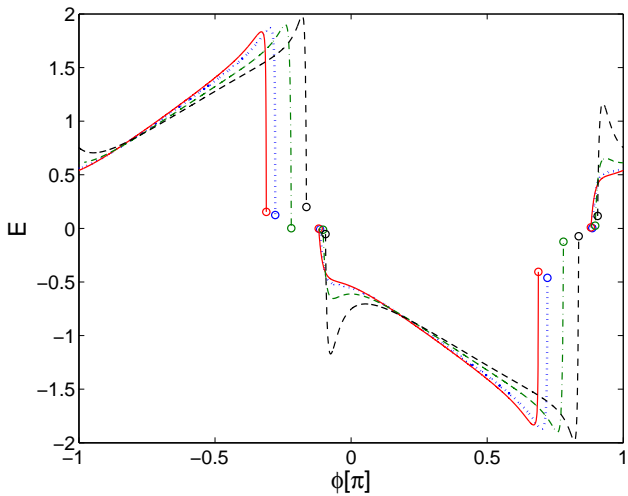}
\caption{Final energy as a function of $\phi$  for a member of a deep penetrating
 {\it disrupted} binary, with a given initial separation $a$. 
Energy is in unit of $(Gm_1m_2/a)(M/m)^{1/3}$ and the phase in unit of $\pi$.
The red solid line is for a binary in radial orbit. Results for
the parabolic formalism are shown for prograde orbits with $D=10^{-3}$ (blue dotted),
 $10^{-2}$ (green dashed-dotted) 
 $3\times10^{-2}$ (black dashed). The boundaries of the intervals in $\phi$
where disruption of the binary occurs are marked with circles.
The phase range is sampled with 6000 equally spaced grid points.
\label{fig:phased2pi}}
\end{figure}

In Figure~\ref{fig:ene_contours}, we show the contour plots for the
final
energy, at $t\gg 1$ of a member of a binary in parabolic orbit that faces either capture by the
black hole ($E={\rm constant}<0$) or ejection ($E={\rm constant}>0$)
from the three-body system. The flat amaranth region is where binaries
are {\em not} disrupted. For a binary approching the black hole on a
prograde orbit (upper panel), a rather shallow penetration factor of
$D=2.1$ is already sufficient to be torn apart by
tidal forces, while no disruption occurs for a retrogarde orbit (lower
panel) with $D \geq 0.44$. 
We chose to plot the lower panel with
reversed axises to emphasize how they tend towards the same
energy distribution  for $D \rightarrow 0$. The limiting distribution depends only on $\phi$ and it is given, of course, by 
the energy plot obtained for radial orbits (Fig.~\ref{fig:phased2pi}).
 Note again that in the radial limit
there is a finite region ($-0.31 \pi< \phi < -0.12 \pi$ and $0.69 \pi < \phi < 0.88 \pi$) where binaries survive.    


\begin{figure}
\plotone{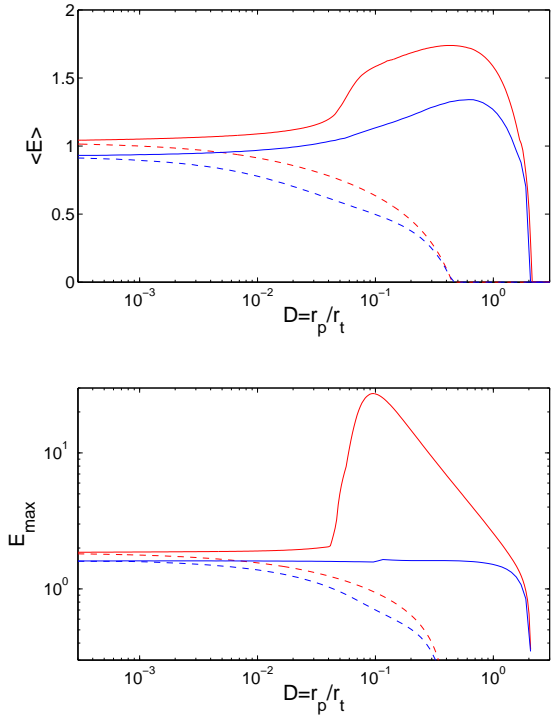}
\epsscale{1.35}\caption{ 
{\it top panel:} Ejection energy averaged over binary phase 
as a function of the penetration factor.
The solid (dashed) line is for prograde (retrograde) orbits. 
The energies for a binary with a given $a$ are plotted as red lines,
in unit of $(Gm_1m_2/a)(M/m)^{1/3}$. The energies binaries
with a given $R_{\rm min}$ are plotted as blue lines, in unit of
$(Gm_1m_2/R_{\rm min})(M/m)^{1/3}$.
{\it bottom panel:} As upper panel, but for the
maximum ejection energy for a given $D$, as a function of $D$. 
The average and maximum values are evaluated for the
absolute value of the energy $|E|$.}
\label{fig:emin_emax}
\end{figure}

Let's now consider the pattern of the energy contours (Figures~\ref{fig:ene_contours} and
~\ref{fig:phased2pi}). For the same
penetration factor, it  shows how a change of $\pi$ in the effective
phase translates into an energy which is equal in absolute value but 
opposite in sign. Therefore, as noted before, a uniform distribution in $\phi$
implies that, when the binary is disrupted, each body has $50\%$ chance
of being ejected. 
Another noticeable feature is the presence and position
of steep narrow peaks next to large plateaux. For prograde orbits,
peaks of $|E|\simeq 27$ are attained {\em not} in the radial limit  
but for a finite penetration factor $D\simeq 10^{-1}$
(see also solid red curve in the lower panel of Fig.~\ref{fig:emin_emax}).  
On the other hand, for retrograde orbits, peaks --or rather ``hills''--
of a more modest maximum energy of $|E| \simeq 1.8$ emerge gradually as
$D$ gets smaller (see red dashed curve, same panel).
Beside the maximal ejection energy, 
we also plot in Fig.~\ref{fig:emin_emax}
the ejection energy averaged over the binary phase, as a function of $D$ (upper panel).
In both plots, a rather large peak is present only for prograde orbits (red solid
curves), and it is situated between $10^{-1} < D <1$. 

A brief summary with the quantitative results of this section follows.
\begin{itemize}
\item The largest $D$ for which there is disruption is 
$D=2.1$ for prograde orbits and $D=0.44$ for retrograde orbits.
\item The maximum energy is 
$27.3$ for prograde orbits and $1.8$ for retrograde orbits.
\item The average energy is
      1.46 for prograde orbits, 0.50 for retrograde orbits,  1.36 for
      prograde and retrograde orbits together. In averaging, we have assumed that $D$,
      or equivalently $r_{\rm p}$, is uniformly distributed. 
\item The highest chance for disruption is for prograde orbits with $D=0.15$.
\end{itemize}

\section{Finite Stellar Sizes}
\label{sec:rmin}

Up until now, we have ignored the finite size of the stars. However, 
the highest energies or ejection velocities are obtained when stars, 
under the action of the black hole tides, get closer to each other before the binary dissolves. 
Specifically, note the sharp peaks in Fig.~\ref{fig:emin_emax} (lower panel)
with exceptional high final energy. Inspection of these orbits,
reveals that the minimal separation that these binaries attain over their evolution 
is more than ten times smaller than their initial separation. Those orbits, however, are not always
physical: if two stars start as almost contact binaries and contract further,
 they may collide and merge or tidally disrupt each other.
We therefore consider in the following a given minimum separation $R_{\rm min}$ --
about the sum of the radii of the two stars-- below which a binary cannot shrink. 

The parabolic and radial formalisms discussed above provide the relative trajectory
of the two stars, but the dimensions are arbitrary and can be rescaled. For each of these trajectories, we can find
the minimum dimensionless distance $\sqrt{x^2+y^2+z^2}|_{\rm min}$  over the whole orbit,
and scale it to be equal to $R_{\rm min}$. 
This amounts to deriving the initial binary separation results in an orbits
whose minimal separation is $R_{\min}$:

$a=  R_{\rm min}/D \sqrt{x^2+y^2+z^2}|_{\rm min}$ in the parabolic formalism, and
$a= R_{\rm min}/\sqrt{x^2+y^2+z^2}|_{\rm min}$
in the radial formalism.

Assuming this $a$ as our initial condition, the energy is given by
\begin{equation}
E_1=-E_2=-{Gm_1m_2\over R_{\rm min}} \left( M \over m \right)^{1/3}  
\sqrt{x^2+y^2+z^2}|_{\rm min} \ 
\end{equation}
\[
\times \left[  {(1+\cos f)^2 \over 4}(x\cos f+y\sin f) + \frac{-\sin f \dot x +(1+\cos f) \dot y }{\sqrt 2}\right],
\]
or by
\begin{equation}
E_1=-E_2=-{Gm_1m_2\over R_{\rm min}} \left( M \over m \right)^{1/3}  
\sqrt{x^2+y^2+z^2}|_{\rm min} \ 
\end{equation}
\[
\times \left(2 \over 9 \right)^{2/3} |t|^{-4/3} \left( x +3 t \dot x  \right),
\]
in our parabolic and radial formalism, respectively. 

\begin{figure}
\epsscale{1.25}
\plotone{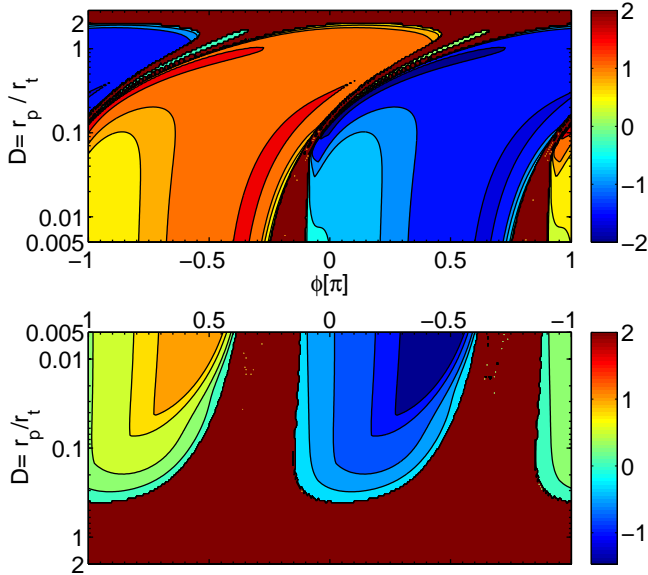}
\caption{As Fig.~\ref{fig:ene_contours} but for a given minimal distance $R_{\rm min}$. 
Here, energy is in unit of $(Gm_1m_2/R_{\rm min})(M/m)^{1/3}$ and the phase in unit of $\pi$.
Note the energy scale, from $-2$ to $2$, a factor of 5 smaller than the energy
scale needed for prograde orbits with a fixed $a$ (Figure~\ref{fig:ene_contours}, upper panel).
\label{fig:enermin_contours}}
\end{figure}

\begin{figure}
\epsscale{1.15}
\plotone{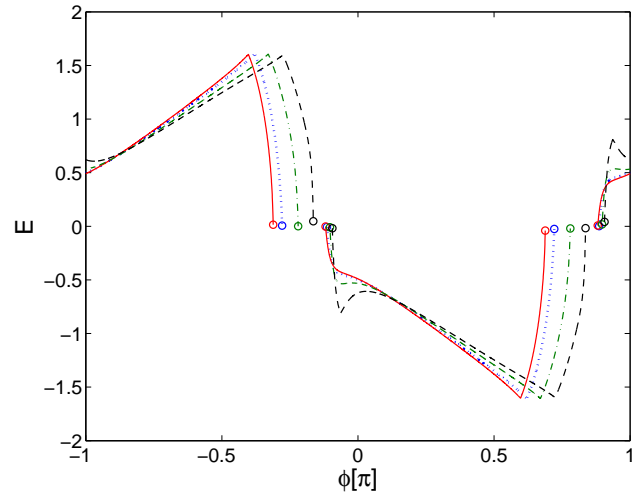}
\caption{As Fig.~\ref{fig:phased2pi} but for a given minimal distance $R_{\rm min}$. 
Energy is in unit of $(Gm_1m_2/R_{\rm min})(M/m)^{1/3}$ and the phase in unit of $\pi$.
\label{fig:ene_rmin_deep}}
\end{figure}

Our analytic approach, therefore, allows us to recast our results in
terms of a given minimal distance between the stars, (Figs.~\ref{fig:enermin_contours} and~\ref{fig:ene_rmin_deep}),
rather than  a given initial separation, (Figs.~\ref{fig:ene_contours} and~\ref{fig:phased2pi}).
In comparison, the main notable feature is that
the scale of Fig.~\ref{fig:enermin_contours} 
is much narrower, i.e.,
 the energy is more sensitive to $R_{\rm min}$ than to $a$. 
This is shown more clearly in Fig.~\ref{fig:emin_emax}, where the
mean and especially the maximal energy for a fixed $R_{\rm min}$ are flatter function of $D$. 
The maximal energies in units of $\left(Gm_1m_2/R_{min}\right)(M/m)^{1/3}$ are:
\begin{itemize}
\item $1.64$ for prograde orbits (obtained for
      $D=0.12$) and $1.60$ for retrograde orbits (obtained for $D=0$).
\end{itemize}


\section{Ejection velocities}
\begin{figure}
\epsscale{1.3}
\plotone{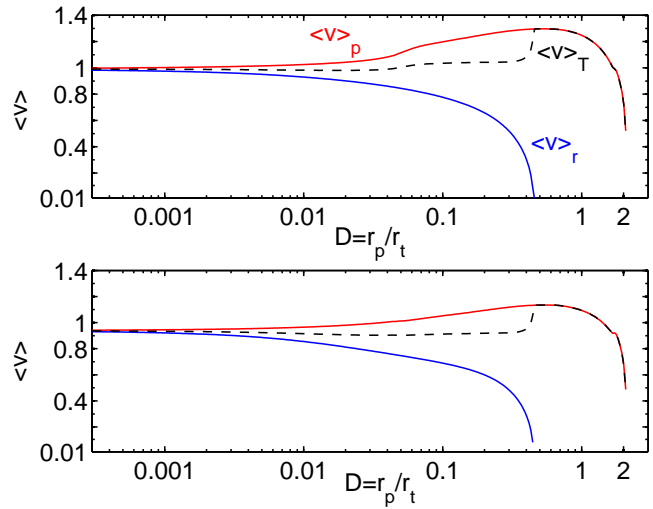}
\caption{The ejection velocity averaged uniformely over phase as a
 function of the penetration factor. The average is over 3000 $\phi$ equally spaced.
 {\it upper panel:} The ejection
 velocity for a given $a$ in unit of $\sqrt{2 G m_{\rm 2}/a}\,(M/m)^{1/6}$. 
 We plot it for
 prograde orbits only (red solid line, $\left<V\right>_{\rm p}$), for retrograde orbits only
 (blue solid line, $\left<V\right>_{\rm r}$) and the weighted average over these two orbital directions
 (black dashed line, $\left<V\right>_{\rm T}$). See text for details. {\it Lower panel:} as upper panel, but for a given
 minimum possible distance $R_{\rm min}$ between stars: here $\left< V \right>$ is in unit of
  $\sqrt{2 G m_{2}/R_{\rm min}}\,(M/m)^{1/6}$. }
\label{fig:vmean_comp}
\end{figure}

In the quest for HVSs the observed quantities are the radial velocity and
the star mass. Neglecting the Galactic potential, an ejected star
($E_1>0$) with mass $m_1$ has a velocity at infinity of $v_1 =\sqrt{2E_1/ m_1}$.
A comprehensive study of velocity and mass distributions, under 
various assumptions for the binary population is deferred to a subsequent paper. 
Here, we calculate the mean velocity for a given binary, assuming that the binary
plane is the same as the orbital plane around the black hole and that  
the binary phase is uniformly distributed. The results are shown in
Fig.~\ref{fig:vmean_comp} (upper panel),  
as a fucntion of the penetration factor.
If we call $\left<v\right>_{\rm p}$, the average velocity
for a prograde orbit (red solid line) and $\left<v\right>_{\rm r}$ that for
a retrograde orbit (blue solid line), then the total average is
$\left<v\right>_{\rm tot}= n_{\rm p}\left<v\right>_{\rm p}+n_{\rm r}\left<v\right>_{\rm r}$
(black-dashed line), where $n_{\rm r}$ and  $n_{\rm p}$ are the fraction of ejected stars 
that originate from retrograde  and prograde binaries. 
Clearly there is no gain in deeper impacts, where the mean saturates at 
$\sim 0.99 \sqrt{2 G m_{\rm 2}/a}\,(M/m)^{1/6}$. Higher velocity --but only by $30\%$-- 
can be attained closer to the tidal radius.
We also calculate similar means, 
but for binaries that, under the black hole tides, shrink
to the minimum relative distance $R_{\rm min}$ (see \S\ref{sec:rmin}). 
Harder binaries -- which would be responsible for the higher mean velocities  --
would now coalesce instead. This results
in an even flatter mean velocity as a function of $D$, which tends to 
$\sim 0.94 \sqrt{2 G m_{2}/R_{\rm min}}\,(M/m)^{1/6}$ for $D \rightarrow 0$,
 (Fig.~\ref{fig:vmean_comp} lower panel).

\section{Discussion and Conclusions}

The ultimate goal of our work is to statistically caracterize
the population of stars originated from tidally disrupted binaries, and
compare it with observations of hyper velocity stars.

To this purpose, we derive in this paper the equations of motion and energy for a member
of a close binary, which suffers an encounter with a third far more massive body. 
For the general case, we assume a parabolic orbit for an effective center of mass
of the binary (\S\ref{sec:parabolic}).
This is in contrast with previous works, \citep[e.g.][]{BKG+06} who considered binaries
on hyperbolic or elliptical orbits.
However, our assumption is justified since the orbits of binaries that are candidates  to
produce hyper velocity stars are very eccentric. The periapsis passage of a very
eccentric orbit could be modeled by a parabolic orbit to a good approximation. This is in 
agreement with the finding of \cite{BKG+06} that the initial binary velocity  is 
of little influence on the final outcome.
Nevertheless, a binary on a hyperbolic orbit has a total positive energy, and 
allows for a new disruption channel, in which both stars are ejected.  
This channel is relatively rare, since the energy of the binary
is typically small compared to the typical ejection energies. 
Moreover, such double ejections could not lead to hypervelocity stars,
since the energy is limited to the original-- small-- positive energy of the binary.
Our formalism with zero total energy (except the small negative binary binding energy),
does not allow for double ejections.

A parabolic trajectory for the binary approaches a radial one for a very close encounter with the massive body.
This observation leads us to explicitly adopt a radial orbit with zero energy,
in order to follow the limiting case of a deep penetrator (\S\ref{sec:radial}). This simpler set of equations allow us
to easily trace a close encounter, that otherwise would require high accuracy
when calculated with a full three-body code.

Our formalism can be applied quite generally to explore the fate
of a binary with arbitrary orbital parameters. 
However, in this paper, we only focused on results for
circular coplanar binaries. The inclination of the binary, 
as well as its eccentricity, is
expected to affect our results quantitatively.
We reserve such a study to a forthcoming paper.
Nevertheless, we can already reach some conclusions and note 
quantitative differences with results found in literature.

The main feature of HVSs is of course their unusually high radial velocity.
For equal mass stars, $m_1=m_2=m_{1,2}$, the expression $Gm_1m_2/R_{\rm min}$ 
is simply $m_{1,2}v_{\rm esc}^2/4$, where $v_{\rm esc}$ is the escape
velocity from the surface of the stars. Since
the maximal ejection energy is
$1.6 ~Gm_1m_2/R_{\rm min}(M/m)^{1/3}$ (Fig.~\ref{fig:emin_emax}),
we derive for the Galactic black hole a corresponding velocity of $0.9 ~v_{\rm esc}(M/m)^{1/6}\cong 9 \,v_{\rm esc}$.
The escape velocity for main sequence stars is about 600-800~km/s in the mass range of 
$1-10 \,M_\odot$. Therefore velocities can be as large as 
 $\sim 7000$~km/s even for binaries limited to the plane. 
Scaling for the different masses (and thus $v_{\rm esc}$) assumed by \cite{Hil88} and by \cite{BKG+06},
we find that a velocity of $0.9 ~v_{\rm esc}(M/m)^{1/6}$
is respectively a factor of $\sim 1.2$ and $\sim 1.8$ higher than their maximum value of $4000$ Km/s. 
Even higher velocities can be acheived 
when the binary mass ratio is large and the {\it lighter} star is ejected.
Then, the ejected star travels with maximal velocity of $1.3 \,v_{\rm esc}(M/m)^{1/6}$,
which is around $\sim 10,000$~km/s for our Galactic center black hole where $M/m \sim 10^6$.

For a comparison with observations, it is also important to determine the ejection probability.
\cite{BKG+06} find that the probability for disruption as a function
of the penetration factor goes roughly as $P_{\rm ej} \approx
1-D/2.2$ . The maximum $D$ for disruption is therefore
very similar to ours, $D_{\rm max} \sim 2.1$
(see Fig.~\ref{fig:vmean_comp}). In contrast, we remark 
a quantitatively different behaviour of the probability function for 
binaries that approach closely the black hole, $D\rightarrow 0$.
Specifically, their $P_{\rm ej} \rightarrow 1$ 
implies that all binaries in this limit
are disrupted. This may look intuitively sound since binaries
that penetrate so deeply experience very strong tides. However, as we stressed
in \S\ref{sec:radial}, the stars in such binaries separate, but approach each other again
on the way back from the black hole. In the planar case we find
that $\sim 20\%$ of the deeply penetrating binaries survive (Fig.~\ref{fig:dis_f}). 
Even when taking a uniform distribution for the inclination of the binary plane into account,
the percentage of survival is still $10\%$  for close encounters.
In the planar case,
Figure~\ref{fig:dis_f} indicates that the disruption probability is, in fact, not
monotonic in $D$: for prograde orbits, it peaks at $D=0.15$  and
it is almost unity ($98.0-98.8\%$) for $0.06 < D < 0.3$.
 
For large mass ratios $m_1 / m_2 > 10$, \cite{BKG+06} find that the
heavier has consistently more chance to be
ejected. We showed here that the probability is 50\%.
This result is not limited to zero inclination of eccentricity.
This fact together with the rarity of a very massive star somewhat weakens the claim
that the star SO-2 was created
by a disruption of a binary, in which a $60\, M_\odot$ companion was ejected \citep{GQ03}.
This conclusion was based on the observed orbital parameters of SO-2. 
However, the timescale to significantly change its shor periapsis distance may quite short, and a careful study 
of the dynamical processes in the inner regions of the Galactic Center is needed to asses it.


\acknowledgments 
This research was partially supported by an ERC grant, 
a Packard Fellowship and a HEFCE PR fellowship. 
We thank Ehud Nakar and Peter Goldreich for helpful discussions.

\bibliographystyle{apj}
\bibliography{hvs}

\end{document}